\documentclass[twocolumn,preprintnumbers,amsmath,amssymb]{revtex4}

\newcommand{\commentoutA}[1]{}

\usepackage{graphicx}% Include figure files
\usepackage{dcolumn}% Align table columns on decimal point
\usepackage{bm}% bold math
\begin{document}

\preprint{LA-UR 12-23991}

\title{Fast method for quantum mechanical molecular dynamics}

\author{Anders M. N. Niklasson\footnote{Corresponding Author Email: amn@lanl.gov}}
\author{Marc J. Cawkwell} 
\affiliation{Theoretical Division, Los Alamos National Laboratory, Los Alamos, New Mexico 87545}

\date{\today}

%\pacs{71.15.Pd,31.15.Qg, 31.15.Ew}% PACS, the Physics and Astronomy
                             % Classification Scheme.
\begin{abstract}
As the processing power available for scientific computing grows, first principles 
Born-Oppenheimer molecular dynamics simulations are becoming increasingly popular 
for the study of a wide range of problems in materials science, chemistry and biology. 
Nevertheless, the computational cost of Born-Oppenheimer molecular dynamics 
still remains prohibitively large for many potential applications.
Here we show how to avoid a major computational bottleneck:
the self-consistent-field optimization prior to the force calculations.
The optimization-free quantum mechanical molecular dynamics method gives
trajectories that are almost indistinguishable from an ``exact'' microcanonical 
Born-Oppenheimer molecular dynamics simulation even when low pre-factor linear scaling sparse 
matrix algebra is used.  Our findings show that the computational gap between
classical and quantum mechanical molecular dynamics simulations can be significantly reduced.
\end{abstract}

\maketitle

\section{Introduction}

The past three decades have witnessed a dramatic increase in the use of the molecular dynamics 
simulation method \cite{MKarplus02,DMarx00}. While it is unquestionably a 
powerful and widely used tool, its ability to calculate physical properties 
is limited by the quality and the computational complexity of the interatomic potentials. 
Among computationally tractable models, the most accurate are 
explicitly quantum mechanical with interatomic forces calculated 
on-the-fly using a nuclear potential energy surface that is determined 
by the electronic ground state within the Born-Oppenheimer approximation \cite{MKarplus73,CLeforestier78,DMarx00}.  
In Hartree-Fock \cite{Roothaan,RMcWeeny60} or density functional theory \cite{hohen,WKohn65,RParr89,RMDreizler90},
the electronic ground-state density is given through a self-consistent-field (SCF)
optimization procedure, which involves iterative mixed solutions of the single-particle
eigenvalue equations and accounts for details in the charge distribution.   
Since the interatomic forces are sensitive to the electrostatic potential \cite{RPFeynman39}, 
molecular dynamics simulations are often of poor quality without a high degree
of self-consistent-field convergence.  This is unfortunate since the iterative 
self-consistent-field procedure is computationally expensive and in practice always approximate.

Recently there have been efforts to reduce 
the computational cost of the self-consistent-field optimization without causing any significant deviation
from ``exact'' Born-Oppenheimer molecular dynamics simulations \cite{PPulay04,ANiklasson06,TDKuhne07}.
In this article we go one step further, and in analogy to time-dependent techniques
such as Ehrenfest molecular dynamics \cite{PEhrenfest27,JAlonso08,JJakowski09} or the
Car-Parrinello method \cite{RCar85,DMarx00,MTuckerman02,BHartke92,HBSchlegel01,JHerbert04,BKirchner12},
we show how the electronic ground state optimization can be circumvented fully without any noticeable
reduction in accuracy in comparison to ``exact'' Born-Oppenheimer molecular dynamics. 

Our optimization-free dynamics is based on a reformulation of extended Lagrangian Born-Oppenheimer
molecular dynamics \cite{ANiklasson08} in the limit of vanishing self-consistent-field optimization.
The method is presented within a general free energy formulation that is valid also at 
finite electronic temperatures and should be applicable to a broad class of materials.  
In addition to the removal of the costly self-consistent-field optimization we also demonstrate
compatibility with low pre-factor linear scaling electronic structure theory \cite{SGoedecker99,DBowler11,ANiklasson02}. 
The combined scheme provides a very efficient, energy conserving, low-complexity method for performing accurate 
quantum molecular dynamics simulations.

\section{Fast Quantum Mechanical Molecular Dynamics}

\subsection{Born-Oppenheimer molecular dynamics}

Born-Oppenheimer molecular dynamics based on density functional theory can be described by the Lagrangian
\begin{equation}\label{BOMD}
{\cal L}^{\rm BO}({\bf R}, {\bf \dot R}) = \frac{1}{2}\sum_k M_k{\dot R}_k^2 - U[{\bf R};\rho],
\end{equation}
where the potential energy,
\begin{equation}\label{DFT_POT}
\begin{array}{l} {\displaystyle U[{\bf R};\rho] = 2\sum_{i \in \rm occ} \varepsilon_i
- \frac{1}{2} \iint \frac{\rho({\bf r})\rho({\bf r'})}{|{\bf r'-r}|} d{\bf r'}d{\bf r}} \\
~~~~~ {\displaystyle- \int V_{\rm xc}[\rho]\rho({\bf r}) d{\bf r} + E_{\rm xc}[\rho] + E_{\rm zz}[{\bf R}],}
\end{array}
\end{equation}
is calculated at the self-consistent electronic ground state density, $\rho({\bf r})$, 
for the nuclear configuration ${\bf R} = \{R_k\}$ \cite{RParr89,RMDreizler90}.
Here, $\varepsilon_i$ are the (doubly) occupied eigenvalues of the effective single-particle Kohn-Sham Hamiltonian,
\begin{equation}\label{HF}
H[\rho] = -\frac{1}{2} \nabla^2 + V_{\rm n}({\bf R},{\bf r}) + \int \frac{\rho({\bf r'})}{|{\bf r'-r}|} d{\bf r'}
+  V_{\rm xc}[\rho],
\end{equation}
where $V_{\rm xc}[\rho]$ is the exchange correlation potential, $V_{\rm n}({\bf R},{\bf r})$ the external (nuclear) potential, and
$-\frac{1}{2}\nabla^2$ the kinetic energy operator. $E_{\rm zz}[{\bf R}]$ is the
electrostatic ion-ion repulsion and $E_{\rm xc}[\rho]$ the exchange correlation energy.

If the electron density deviates from the ground state density 
$\rho$ by some small amount $\delta \rho$,
the error in the potential energy is essentially of the order $\delta \rho^2$, depending on
the particular formulation used for calculating $U[{\bf R};\rho+\delta \rho]$ \cite{JHarris85,APSutton88,WMCFoulkes89}.
However, since the Hellmann-Feynman theorem is valid only at the ground state density, we do not 
have a simple expression for the forces that avoids calculating derivatives of the electronic density, 
$\partial (\rho+\delta \rho) / \partial R_k$. In practical calculations, the accuracy 
of the potential energy can therefore not be expected to hold also for the forces and
a high degree of self-consistent-field convergence is therefore typically required.

\subsection{Extended Lagrangian molecular dynamics}

Here we outline how we can circumvent the self-consistent-field procedure in Born-Oppenheimer molecular dynamics.
Instead of recalculating the ground state density before each force evaluation with an iterative optimization procedure, 
the idea here is to use an auxiliary density $n({\bf r})$, as in extended Lagrangian Born-Oppenheimer 
molecular dynamics \cite{ANiklasson08,ANiklasson09,PSteneteg10,GZheng11,ANiklasson11}, which evolves through a harmonic 
oscillator centered around the ground state density $\rho({\bf r})$. 
Based on a general free energy formulation of extended Lagrangian Born-Oppenheimer molecular dynamics \cite{ANiklasson11}
in the limit of vanishing self-consistent-field optimization, we define the extended Lagrangian:
\begin{equation}\label{X}\begin{array}{l}
{\displaystyle {\cal L}({\bf R}, {\bf \dot R},n,{\dot n}) = \frac{1}{2}\sum_k M_k{\dot R}_k^2 
-  {\cal U}[{\bf R};n] + T_e{\cal S}[{\bf R};n] } \\ 
{\displaystyle + \frac{1}{2} \mu \int {\dot n}({\bf r})^2d{\bf r}
- \frac{1}{2} \mu \omega^2\int \left(\rho({\bf r})-n({\bf r})\right)^2d{\bf r}}.
\end{array}
\end{equation}
While the potential and entropy terms, ${\cal U}$ and ${\cal S}$, are well defined 
at the ground state density \cite{RParr89}, i.e. when $n=\rho$, there are several different options
when $n$ deviates from $\rho$, e.g. the Harris-Foulkes functional \cite{JHarris85,APSutton88,WMCFoulkes89}.
In a more general case, the potential energy and entropy term 
may therefore also be determined by $n({\bf r})$ implicitly through an additional function $\sigma[n({\bf r})]$,
i.e.\ ${\cal U}[{\bf R};n] \equiv {\cal U}[{\bf R};n,\sigma[n]]$ and ${\cal S}[{\bf R};n] \equiv {\cal S}[{\bf R};n,\sigma[n]]$. Here
$\sigma[n({\bf r})]$ is a temperature dependent density given from the diagonal part of the real-space representation
of the (doubly occupied) density matrix, which is given through a Fermi-operator expansion \cite{RParr89}
of the effective single-particle Hamiltonian, $H[n]$, i.e.
\begin{equation}\label{EL_4c}
\sigma({\bf r}) \equiv \sigma[n({\bf r})] = 2\left.{\left(e^{\beta(H[n]-\mu_0 I)}+1\right)^{-1}}\right\rvert_{\bf r = r'}.
\end{equation}
At zero electronic temperature the Fermi-operator expansion corresponds to a step function
with the step formed at the chemical potential, $\mu_0$.
In our Lagrangian above, $\mu$ and $\omega$ are fictitious mass and frequency parameters of the harmonic oscillator and $\beta$ is
the inverse electronic temperature, i.\ e.\ $\beta = 1/(k_{\rm B}T_e)$.
The purpose of the entropy-like term ${\cal S}[{\bf R};n]$ is here
to make the derived forces of our dynamics variationally correct for a given 
entropy-independent density, $n({\bf r})$, at any electronic temperature. 
This approach is different from the regular formulation where the density 
is determined by the entropy through the minimization of the electronic free energy functional
\cite{RParr89,MWeinert92,RWentzcovitch92,ANiklasson08b}.

\subsubsection{Equations of motion}

The molecular trajectories corresponding to the extended free energy Lagrangian ${\cal L}$ in Eq.\ (\ref{X})
are determined by the Euler-Lagrange equations of motion,
\begin{equation}\label{0a}\begin{array}{l}
{\displaystyle M_k{\ddot R}_k = \left.{-\frac{\partial {\cal U}[{\bf R};n]}{\partial R_k}}\right\rvert_{n}
+ T_e\left.{\frac{\partial {\cal S}[{\bf R};n]}{\partial R_k}}\right\rvert_{n} } \\
~~\\
{\displaystyle - \left.{\frac{\mu \omega^2 }{2}\frac{\partial}{\partial R_k} \int \left(\rho({\bf r})-n({\bf r})\right)^2d{\bf r}}\right\rvert_{n}},\\
\end{array}
\end{equation}
and
\begin{equation}\label{EL_1b}\begin{array}{l}
{\displaystyle \mu {\ddot n}({\bf r}) = \mu \omega^2 \left(\rho({\bf r})-n({\bf r})\right)} \\
~~\\
{\displaystyle - \left.{\frac{\delta {\cal U}[{\bf R};n]}{\delta n}}\right\rvert_{\bf R} 
+ T_e\left.{\frac{\delta {\cal S}[{\bf R};n]}{\delta n}}\right\rvert_{\bf R}},
\end{array}
\end{equation}
where the partial derivatives are taken with respect to constant density, $n$, or coordinates, ${\bf R}$.
The limit $\mu \rightarrow 0$ gives us the equations of motion of our extended Lagrangian dynamics,
\begin{equation}\label{EL_3a}
M_k{\ddot R}_k = -\left.{\frac{\partial {\cal U}[{\bf R};n]}{\partial R_k}}\right\rvert_{n} + T_e\left.{\frac{\partial {\cal S}[{\bf R};n]}{\partial R_k}}\right\rvert_{n}
\end{equation}
\begin{equation}\label{EL_3b}
{\ddot n}({\bf r}) = \omega^2 \big(\rho({\bf r})-n({\bf r})\big),
\end{equation}
where we have defined ${\cal S}[{\bf R};n]$ such that 
\begin{equation}\label{Def_Ent}
\left.{\frac{\delta {\cal U}[{\bf R};n]}{\delta n}}\right\rvert_{\bf R}  = T_e\left.{\frac{\delta {\cal S}[{\bf R};n]}{\delta n}}\right\rvert_{\bf R}.
\end{equation}
As is shown in the Appendix (Sec.\ \ref{HF_appendix}), 
the corresponding property for $\partial {\cal S}/\partial R_k$ is also of importance for the 
calculation of the Pulay force in Eq.\ (\ref{EL_3a}).
Notice that these equations still require a full self-consistent field optimization, 
since the auxiliary density $n({\bf r})$ evolves around the ground state density $\rho({\bf r})$.

Since the nuclear degrees of freedom do not depend on the mass parameter $\mu$ in Eqs.\ (\ref{EL_3a}) and (\ref{EL_3b}), 
the total free energy,
\begin{equation}\label{Etot}
E_{\rm tot} = \frac{1}{2}\sum_k M_k{\dot R}_k^2 + {\cal U}[{\bf R};n] - T_e{\cal S}[{\bf R};n],
\end{equation}
is a constant of motion in the limit of vanishing $\mu$. 
Moreover, if $E_{\rm tot}$ is close to the exact ground state free energy
for approximate densities $n({\bf r})$, we can also expect that the forces of 
the extended Lagrangian dynamics should be accurate. 

The forces in Eq.\ (\ref{EL_3a}) are calculated at the approximate,
unrelaxed, density $n({\bf r})$ using a Hellmann-Feynman-like
expression, where the partial derivatives are taken with respect to a constant density $n({\bf r})$.
This is possible only because $n({\bf r})$ appears as an independent dynamical variable.
In general, as mentioned above, this can not be assumed, since the Hellmann-Feynman
force expression is formally applicable only at the ground density.
A more detailed derivation of explicit force expressions is given in the Appendix.

\subsubsection{Entropy contribution}

Depending on the particular functional form chosen for the potential energy term, ${\cal U}({\bf R};n)$,
we may not have access to a simple explicit expression of ${\cal S}[{\bf R};n]$ that fulfills Eq.\ (\ref{Def_Ent}). 
In this case an approximate entropy term has to be used.
This has no effect on the dynamics in Eqs.\ (\ref{EL_3a}) and (\ref{EL_3b}), 
since the forces remain exact by definition. An approximation of the entropy term therefore
only affects the estimate of the constant of motion, $E_{\rm tot}$, in Eq.\ (\ref{Etot}). 

We have found that the regular expression for the electronic entropy \cite{RParr89},
\begin{equation}\label{Entropy}
{\cal S}[{\bf R};n] =  -2k_{\rm B} \sum_i \left\{f_i \ln(f_i) + (1-f_i)\ln(1-f_i)\right\},
\end{equation}
which formally is defined only at the ground state density, i.e. when $n = \rho$, typically
provides a highly accurate approximation also for approximate densities as will be illustrated in the examples below.
Here $f_i$ are the occupation numbers of the states, i.e. the eigenvalues of the density matrix in Eq.\ (\ref{EL_4c}).
These are determined by the Fermi-Dirac distribution of the single-particle eigenvalues $\varepsilon_i$ of the Hamiltonian $H[n]$, i.e.
\begin{equation}
f_i = \left[ e^{\beta(\varepsilon_i - \mu)} + 1\right]^{-1}.
\end{equation}
By comparing the calculation of $E_{\rm tot}$ in Eq.\ (\ref{Etot}) using the approximate entropy term, 
${\cal S}[{\bf R};n]$, in Eq.\ (\ref{Entropy})
to ``exact'', fully optimized, Born-Oppenheimer molecular simulations, we can estimate the accuracy of our dynamics.

\subsection{Fast quantum mechanical molecular dynamics}

As in extended Lagrangian Born-Oppenheimer molecular dynamics, 
the irreversibility of regular Born-Oppenheimer molecular dynamics
that is caused by the self-consistent-field optimization,
can be avoided, since the density $n({\bf r})$ can be integrated using a reversible geometric 
integration algorithm \cite{BLeimkuhler04,ANiklasson08,ANiklasson09,AOdell09}, e.g. the Verlet algorithm as in Eq.\ (\ref{Verl_n}) below.
This prevents the unphysical drift in the energy and phase space of regular Born-Oppenheimer molecular dynamics 
\cite{PPulay04,ANiklasson06,TDKuhne07} and
our dynamics will therefore exhibit long-term stability of the free energy $E_{\rm tot}$ in Eq.\ (\ref{Etot}).

A main problem so far is that we still need to calculate the self-consistent ground state density $\rho({\bf r})$
in the integration of $n({\bf r})$ in Eq.\ (\ref{EL_3b}).
Fortunately, various geometric integrations of the auxiliary density $n({\bf r})$ in  Eq.\ (\ref{EL_3b})
are stable also for approximate ground state density estimates of $\rho({\bf r})$, as long as the approximation of 
$\rho({\bf r})$ is at least infinitesimally closer to the exact ground state compared to $n({\bf r})$. 
Using an integration time step of $\delta t$, this stability holds if
the value of the dimensionless variable $\kappa = \delta t^2 \omega^2$ is chosen to be appropriately small \cite{ANiklasson09,AOdell09,AOdell11}.
For energy functionals that are convex in the vicinity of the ground state density we may therefore
replace $\rho({\bf r})$ in Eq.\ (\ref{EL_3b}) by a linear combination $(1-c)n +c\sigma$ 
\cite{PHDederichs83}, which gives us the approximate equations of motion 
\begin{equation}\label{EL_3ax}
M_k{\ddot R}_k = -\left.{\frac{\partial {\cal U}[{\bf R};n]}{\partial R_k}}\right\rvert_{n} 
+ T_e\left.{\frac{\partial {\cal S}[{\bf R};n]}{\partial R_k} }\right\rvert_{n},
\end{equation}
and
\begin{equation}\label{EL_3bx}
{\ddot n}({\bf r}) = \omega^2 \big(\sigma({\bf r})-n({\bf r})\big),
\end{equation}
where the constant $\omega^2$ has been rescaled by $c$.
The Verlet integration of Eq.\ (\ref{EL_3bx}), including a weak dissipation to avoid an accumulation 
of numerical noise \cite{ANiklasson09,PSteneteg10}, 
\begin{equation}\label{Verl_n}
{\displaystyle n_{t+\delta t} = 2n_{t} - n_{t-\delta t} +  \delta t^2 \omega^2\left(\sigma_t - n_t\right) + \alpha \sum_{k=0}^K c_k n_{t-k\delta t},}
\end{equation}
is therefore stable if a sufficiently small positive value of $\kappa = \delta t^2 \omega^2$
is chosen \cite{ANiklasson09}.  Thus, without any self-consistent-field optimization of $\rho({\bf r})$,
the previously optimized values of $\kappa$ in Ref.\ \cite{ANiklasson09,AOdell09,AOdell11} should be rescaled by a positive factor $\le 1$.
Certain ill behaved (non-convex) functionals with self-consistent-field instabilities \cite{PHDederichs83} can 
not be treated in this framework.

The proposed molecular dynamics as given by Eqs.\ (\ref{EL_3ax}) and (\ref{EL_3bx})
is the central result of this paper. The equations of motion
do not involve any ground state self-consistent-field optimization 
prior to the force evaluations and only one single diagonalization or density matrix construction 
is required in each time step.  
The frequency $\omega$ of the electronic density is well separated from the nuclear vibrational oscillations.
Using a value of $\delta t \omega = {\sqrt \kappa} \approx 1$ and an integration time step $\delta t$,
which is $\sim 1/15$ of the period of the nuclear motion, the frequencies differ by a factor of 5.
As will be demonstrated in the examples below, the scheme 
is also fully compatible with linear scaling electronic structure theory \cite{SGoedecker99,DBowler11}.
This compatibility is crucial in order to simulate large systems.
The removal of the costly ground state optimization, 
in combination with low-complexity linear scaling solvers, 
provide a computationally fast quantum mechanical molecular dynamics (Fast-QMMD) that can match
the fidelity and accuracy of regular Born-Oppenheimer molecular dynamics.

There are several alternative approaches to derive or motivate the equations of motion of
the fast quantum mechanical molecular dynamics, Eqs.\ (\ref{EL_3ax}) and (\ref{EL_3bx}), and
details of the dynamics may vary depending on the choice of the functional form of ${\cal U}({\bf R};n)$. 
However, the particular derivation presented here is the most transparent and general 
approach that we have found so far.

The equations of motion are given in terms of the electron density, 
but they should be generally applicable to a large class of methods, such as Hartree-Fock theory, 
which is analyzed in the Appendix (Sec.\ \ref{HF_appendix}), or plane wave pseudo-potential methods \cite{PSteneteg10}.  
Here we will demonstrate our fast quantum mechanical molecular dynamics scheme using 
self-consistent-charge density functional tight-binding theory \cite{APSutton88,MFinnis98,MElstner98,MFinnis03},
as implemented in the electronic structure code \textsc{latte} \cite{ESanville10}, 
either with an orthogonal or a non-orthogonal representation and 
both at zero and at finite electronic temperatures. With this method we can easily reach 
the time and length scales necessary to establish long-term energy conservation and 
linear scaling of the computational cost.  Details of the computational method and our particular
choices of ${\cal U}({\bf R};n)$ are given in the Appendix.

\begin{figure}[t]
\resizebox*{3.5in}{!}{\includegraphics[angle=00]{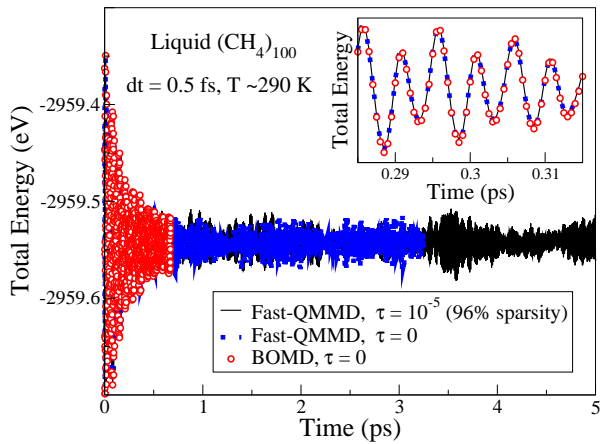}}
\caption{\label{Figure_1}\small
Total energy fluctuations, Eq.\ (\ref{Etot}), using
``exact'' (4 SCF/step) Born-Oppenheimer molecular dynamics (BOMD),
and the fast quantum mechanical molecular dynamics, Eqs.\ (\ref{EL_3ax}) and (\ref{EL_3bx}), (Fast-QMMD), 
with ($\tau > 0$) or without ($\tau = 0$) thresholding applied in 
the low pre-factor linear scaling solver \cite{ANiklasson02}. The simulations were performed with the 
molecular dynamics program \textsc{latte} using self-consistent-charge density functional based
tight-binding theory in an orthogonal formulation at $T_e = 0$, i.e.\
as in Eqs.\ (\ref{EL_Orth_a}), (\ref{EL_Orth_b}) and (\ref{Etot_Orth}).}
\end{figure}

\begin{figure}[t]
\resizebox*{3.5in}{!}{\includegraphics[angle=00]{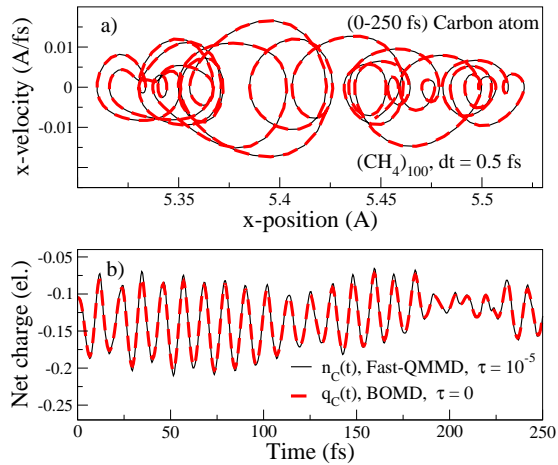}}
\caption{\label{PhSp_C}\small
Panel a) shows the x-plane phase space trajectory of a single carbon atom (C) 
based on an ``exact'' (4 SCF/step) Born-Oppenheimer molecular dynamics (BOMD, dashed line) 
and the fast quantum mechanical molecular dynamics (Fast-QMMD, solid line). 
Panel b) shows the fluctuations in the net auxiliary charge $n_i(t)$ and ground state 
charge $q_i(t)$ for the same carbon atom ($i$=C). The numerical threshold $\tau$
is applied in the linear scaling solver \cite{ANiklasson02}. The simulations were performed with the program 
\textsc{latte} using self-consistent-charge tight-binding theory in an orthogonal 
formulation at zero electronic temperature, i.e.\ as in Eqs.\ (\ref{EL_Orth_a}), (\ref{EL_Orth_b}) and (\ref{Etot_Orth}).}
\end{figure}

\section{Examples}\label{Examples}

\subsection{Orthogonal representation}

\begin{table}[t]
  \centering
  \caption{\protect Wall clock timings of the fast quantum mechanical molecular dynamics (Fast-QMMD) simulations in comparison 
   to Born-Oppenheimer molecular dynamics (BOMD) (4 SCF/step), without ($\tau = 0$) and with ($\tau > 0$) a low pre-factor 
   linear scaling solver for the density matrix \cite{ANiklasson02}
   with threshold tolerance $\tau$. The program (\textsc{latte} in its orthogonal formulation at $T_e = 0$) was executed 
   on a single core of a 2.66 GHz Quad-Core Intel Xeon processor. 
  }\label{Tab_timing}
  \begin{ruledtabular}
  \begin{tabular}{ll}
    Polyethene chain C$_{100}$H$_{202}$  &  Efficiency \\
     \hline
    BOMD ($\tau = 0$) & 7.5 s/step \\
    Fast-QMMD ($\tau = 0$) & 1.5 s/step \\
    Fast-QMMD ($\tau = 10^{-5}$) & 0.61 s/step \\
     \hline
     \hline
    Liquid Methane (CH$_4$)$_{100}$  &  Efficiency\\
     \hline
    BOMD ($\tau = 0$) & 12.5 s/step \\
    Fast-QMMD ($\tau = 0$) & 2.5 s/step \\
    Fast-QMMD ($\tau = 10^{-5}$) & 0.35 s/step \\
  \end{tabular}
  \end{ruledtabular}
\end{table}

Figure \ref{Figure_1} shows the fluctuations in the total energy (kinetic plus potential) using the 
fast quantum mechanical molecular dynamics, Eqs.\ (\ref{EL_3ax}) and (\ref{EL_3bx}), 
as implemented in Eqs.\ (\ref{EL_Orth_a}), (\ref{EL_Orth_b}) and (\ref{Etot_Orth}),
and an ``exact'' Born-Oppenheimer molecular dynamics \cite{ANiklasson08}, for liquid methane (density = 0.422 g/cm$^3$)
at room temperature. The calculations were performed with the \textsc{latte} molecular dynamics program 
using periodic boundary conditions and an integration time step of $\delta t = 0.5$ fs.
Since the molecular system is chaotic, any infinitesimally small deviation will eventually lead to a divergence between
different simulations.
However, even after hundreds of time steps and over 300 fs of simulation time 
the total energy curves are virtually on top of each other as is seen in the inset. 
The same remarkable agreement is seen in Fig.\ \ref{PhSp_C}, which shows the projected phase space of an individual carbon atom
and the fluctuations of its net charge. In this case the C atom was displaced compared to the simulation in Fig.\ \ref{Figure_1} 
to further enhance the charge fluctuations.

\subsection{Linear scaling}

The fast quantum mechanical molecular dynamics scheme is also stable in combination with approximate
linear scaling sparse matrix algebra \cite{SGoedecker99,DBowler11}. 
Using the recursive second order spectral projection method for the construction of the density matrix \cite{ANiklasson02} 
with a numerical threshold, $\tau = 10^{-5}$, below which all elements are set to zero after each individual projection, 
we notice excellent accuracy and stability in Fig. \ref{Figure_1} without any systematic drift in the total energy. 

Despite their high efficiency and low computational pre-factor compared to alternative 
linear scaling electronic structure methods \cite{EHRubensson11}, it has been argued that recursive purification algorithms
are non-variational and therefore incompatible with forces of a conservative system \cite{DBowler11}, which
is necessary for long-term energy conservation. 
As is evident from Figs.\ \ref{Figure_1} and \ref{PhSp_C}, this is not a problem.
The graphs are practically indistinguishable from ``exact'' Born-Oppenheimer molecular dynamics, without
any signs of a systematic drift in the total energy.
The corresponding linear scaling compatibility with microcanonical simulations was recently also demonstrated 
for self-consistent-field-optimized extended Lagrangian Born-Oppenheimer molecular dynamics \cite{MJCawkwell12}.

The gain in speed using the fast quantum mechanical molecular dynamics scheme in comparison to 
Born-Oppenheimer molecular dynamics is illustrated by the wall-clock timings shown in Tab. \ref{Tab_timing}.

\begin{figure}[t]
\resizebox*{3.5in}{!}{\includegraphics[angle=00]{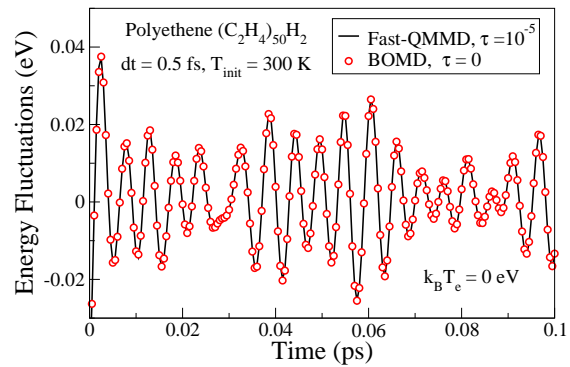}}
\caption{\label{Figure_3}\small
Total energy fluctuations, Eq.\ (\ref{Etot}), using ``exact'' (4 SCF/step) Born-Oppenheimer molecular dynamics (BOMD),
and the fast quantum mechanical molecular dynamics, Eqs.\ (\ref{EL_3ax}) and \ref{EL_3bx}), (Fast-QMMD),
with ($\tau = 10^{-5}$) or without ($\tau = 0$) thresholding applied in
the low pre-factor linear scaling solver \cite{ANiklasson02}. The simulations were performed with the 
molecular dynamics program \textsc{latte} 
in the non-orthogonal formulation at $k_{\rm B}T_e = 0$ eV, 
i.e.\ as implemented in Eqs.\ (\ref{EL_Nonorth_a}), (\ref{EL_Nonorth_b}) 
and (\ref{Etot_NonOrth}) with the entropy term approximated
by ${\cal S} = 0$.}
\end{figure}

\begin{figure}[t]
\resizebox*{3.5in}{!}{\includegraphics[angle=00]{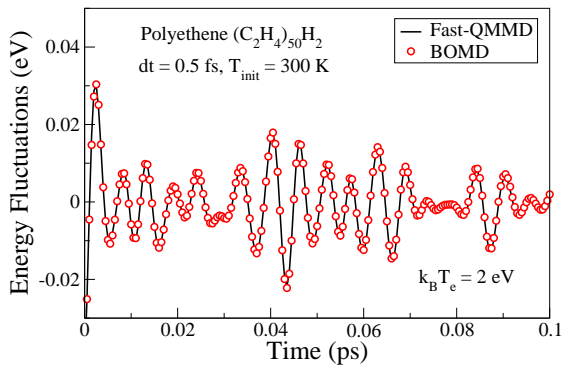}}
\caption{\label{Figure_4}\small
Total free energy fluctuations, Eq.\ (\ref{Etot}), using ``exact'' (4 SCF/step) Born-Oppenheimer molecular dynamics (BOMD),
and the fast quantum mechanical molecular dynamics, Eqs.\ (\ref{EL_3ax}) and (\ref{EL_3bx}), (Fast-QMMD).
The simulations were performed with the molecular dynamics program \textsc{latte} using
the non-orthogonal formulation at an electronic temperature of $k_{\rm B}T_e = 0.5$ eV, i.e. as implemented in
Eqs.\ (\ref{EL_Nonorth_a}), (\ref{EL_Nonorth_b}) and (\ref{Etot_NonOrth}) 
with the entropy term approximated by Eq.\ (\ref{Ent_Approx}).}
\end{figure}

\subsection{Non-orthogonal representation}

For non-orthogonal representations at finite electronic temperatures, 
a Pulay force term and a finite approximate entropy contribution to 
the total free energy have to be included. Figures \ref{Figure_3} and \ref{Figure_4}
illustrate the total energy fluctuations for the fast quantum mechanical molecular dynamics simulations 
of a hydrocarbon chain as implemented in \textsc{latte} using
Eqs.\ (\ref{EL_Nonorth_a}), (\ref{EL_Nonorth_b}) and (\ref{Etot_NonOrth}),
with the approximate entropy term in Eq.\ (\ref{Ent_Approx}).
The electronic temperature of the examples in Figure \ref{Figure_3} is set to zero, $k_{\rm B}T_e = 0$ eV,
and for the examples in Fig. \ref{Figure_4}, $k_{\rm B}T_e = 2$ eV.  In the first time step the initial nuclear temperature, 
$T_{\rm init}$, was set to $300$ K using a Gaussian distribution
of the velocities. Despite the approximation of $\rho$ in Eq.\ (\ref{EL_3bx}) and
the approximate estimate of the entropy contribution to the free energy there is virtually
no difference seen between the fast quantum mechanical and the Born-Oppenheimer molecular dynamics simulations.

As in the orthogonal case, the non-orthogonal
formulation of our fast quantum mechanical molecular dynamics is fully compatible
with linear scaling complexity in the construction of the density matrix at $T_e = 0$ K.
In Fig. \ref{Figure_3} the reduced complexity simulation shows 
no significant deviation from ``exact'' Born-Oppenheimer molecular dynamics.
At finite electronic temperatures, the linear scaling construction of 
the Fermi operator \cite{ANiklasson03,ANiklasson08b} has not yet been implemented.

\subsection{Long-term stability and conservation of the total energy}

To assess the long-term energy conservation and the stability we use a test system comprised of 16 molecules of isocyanic 
acid, HNCO, at a density of 1.14 g cm$^{-3}$. The system was first thermalized to a temperature 
of 300 K over a simulation time of 12.5 ps by the rescaling of the nuclear velocities. 
The simulations used an integration time step, $\delta t$, of 0.25 ps.
The simulations were performed using self-consistent
tight-binding theory \cite{APSutton88,MFinnis98,MElstner98,MFinnis03} with a non-orthogonal basis as implemented in \textsc{latte}, 
using Eqs.\ (\ref{EL_Nonorth_a}), (\ref{EL_Nonorth_b}) and (\ref{Etot_NonOrth})
with the entropy term approximated by Eq.\ (\ref{Ent_Approx}).

\begin{figure}
\includegraphics[width=1.0\columnwidth]{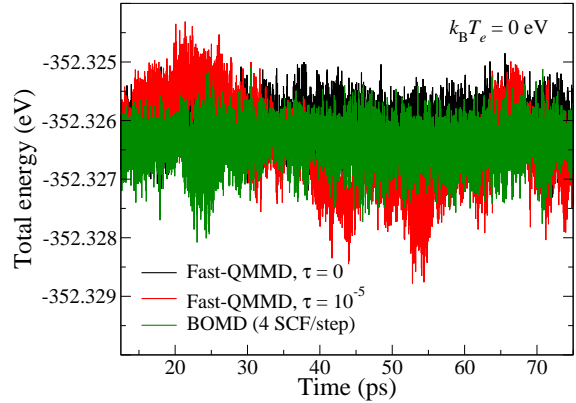}
\caption{Total energy versus time for liquid isocyanic acid with a nuclear temperature of 300 K 
and $T_e = 0$ K computed with ``exact'' Born-Oppenheimer MD and the Fast QMMD method with 
exact and approximate linear scaling density matrix constructions. The numerical threshold $\tau$
is applied in the linear scaling solver \cite{ANiklasson02} below which all matrix elements
are set to zero after each iteration.} \label{zeroT}
\end{figure}

Fast quantum mechanical molecular dynamics  and ``exact'' Born-Oppenheimer molecular dynamics simulations
 with 4 self-consistent field cycles per time step were performed over 250,000 time steps (62.5 ps) with 
$T_e = 0$ K and $k_\text{B}T_e = 0.5$ eV. 
The latter temperature is small with respect to the HOMO-LUMO gap of HNCO, which is about  
6.0 eV, yet the entropy term, Eq.~(\ref{Entropy}) or Eq.\ (\ref{Ent_Approx}), contributes about 0.19 eV to 
the total energy owing to the partial occupation of states in the vicinity of the chemical potential. 
Trajectories computed at $T_e = 0$ K with ``exact'' Born-Oppenheimer molecular dynamics and 
the fast quantum mechanical molecular dynamics method without ($\tau = 0$) and with ($\tau = 10^{-5}$)
linear scaling constructions of the density matrix are presented in Fig.~\ref{zeroT}. 
The standard deviation of the fluctuations of the total energy about its 
mean and an estimate of the level of the systematic drift of the total energies are presented 
in Table \ref{example_summary}. These data show that the fast quantum mechanical molecular dynamics
simulations yield trajectories 
that are effectively indistinguishable from the ``exact'' Born-Oppenheimer trajectories. 
Moreover, as was seen above, the fast quantum mechanical molecular dynamics scheme appears to be fully compatible with linear scaling 
construction of the density matrix and the resulting approximate forces, since this trajectory differs from 
the ``exact'' Born-Oppenheimer molecular dynamics trajectory only by a small-amplitude random-walk of the total energy 
about its mean \cite{MJCawkwell12}. 
The systematic drift in energy is several orders of magnitude smaller than in previous attempts to combine
linear scaling solvers with regular Born-Oppenheimer molecular dynamics \cite{FMauri94,APHorsfield96,ETushida08,FShimojo08}.

\begin{figure}
\includegraphics[width=1.0\columnwidth]{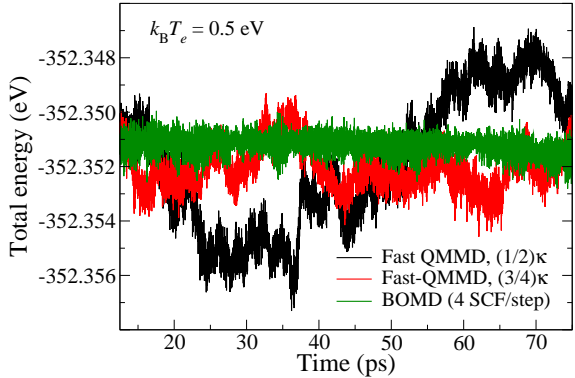}
\caption{Total free energy versus time for liquid isocyanic acid with a nuclear temperature 
of 300 K and $k_\text{B}T_e = 0.5$ eV computed with ``exact'' 
Born-Oppenheimer molecular dynamics (BOMD) and the 
fast quantum mechanical molecular dynamics (Fast-QMMD) method with $\kappa$ rescaled by 3/4
instead of 1/2.} \label{halfeV}
\end{figure}

The trajectories computed with an electronic temperature corresponding to $k_\text{B}T_e = 0.5$ eV
differ qualitatively from those computed with zero electronic temperature. Figure \ref{halfeV} and 
Table \ref{example_summary} show that while the ``exact'' Born-Oppenheimer trajectory 
conserves the free energy to an extremely high tolerance 
over the duration of the simulation, the total free energy in the fast quantum mechanical molecular dynamics simulation 
exhibit random-walk behaviour about the mean value.  
Although the fast quantum mechanical molecular dynamics simulations involve an approximate expression 
for the entropy, we find that this alone cannot account for the level of fluctuations observed. 
Instead, we have found that the rescaling of
the $\kappa$ value in the integration, Eq.\ (\ref{Verl_n}), affects this random-walk. By changing the rescaling
factor to 3/4, instead of 1/2 as in all the other examples, the amplitude of the random walk is significantly reduced.
Nevertheless, the fast quantum mechanical molecular dynamics trajectories 
at finite electronic temperature exhibit systematic drifts in the total energy that are negligible and 
the fluctuations of the total energy about the mean are of the same order as those that arise from 
the application of the approximate linear scaling method at $T_e = 0$ K.

\begin{table}
\caption{Standard deviation, $\sigma$, of the total energy about its mean value
and the upper bound of the systematic drift of the total energy, $E_\text{drift}$, 
computed from ``exact'' Born-Oppenheimer molecular dynamics (BOMD) and 
fast quantum mechanical molecular dynamics (Fast-QMMD) simulations of 
liquid isocyanic acid. The simulation were performed with the \textsc{latte} program 
in the non-orthogonal formulation, i.e. as implemented in 
Eqs.\ (\ref{EL_Nonorth_a}), (\ref{EL_Nonorth_b}) and (\ref{Etot_NonOrth}) 
with the entropy term approximated by Eq.\ (\ref{Ent_Approx}).}\label{example_summary}
\begin{ruledtabular}
\begin{tabular}{llcc}
 $k_\text{B} T_e$& & $\sigma$  & $E_\text{drift}$ \\
 (eV)& & ($\mu$eV) &  ($\mu$eV/atom/ps) \\
 \colrule
& Fast-QMMD ($\tau = 0$) & 0.315 & $5.10 \times 10^{-3}$ \\
0.0 & Fast-QMMD ($\tau = 10^{-5}$) & 0.702 & 0.285 \\
& BOMD (4 SCF/step)  & 0.358 & $9.94 \times 10^{-3}$ \\
\colrule
& Fast-QMMD $(1/2)\kappa$ & 2.47 & 1.43 \\
0.5 & Fast-QMMD $(3/4)\kappa$ & 0.786 & $7.85 \times 10^{-2}$ \\
& BOMD (4 SCF/step) & 0.361 & $8.50 \times 10^{-2}$ \\
\end{tabular}
\end{ruledtabular}
\end{table}

\section{Convergence properties}\label{Conv}

\begin{figure}[t]
\resizebox*{3.5in}{!}{\includegraphics[angle=00]{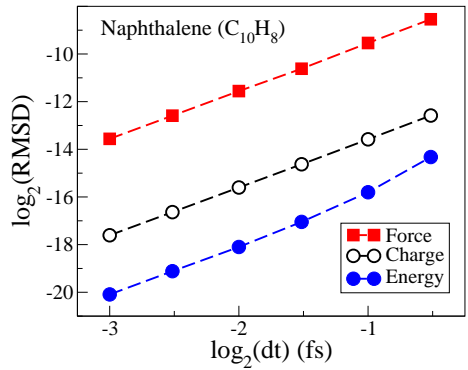}}
\caption{\label{Figure_7}\small
The root mean square deviation (RMSD) between the fast quantum mechanical molecular dynamics, 
Eqs.\ (\ref{EL_3ax})-(\ref{EL_3bx}), and ``exact'' (4 SCF/step) Born-Oppenheimer molecular dynamics, for the nuclear
forces, the net Mulliken charges and the total energy for a Naphthalene molecule at room temperature.
The simulation were performed with the \textsc{latte} molecular dynamics program using
self-consistent-charge tight-binding theory in an orthogonal formulation 
at $T_e = 0$, i.e.\ as implemented in Eqs.\ (\ref{EL_Orth_a}), 
(\ref{EL_Orth_b}) and (\ref{Etot_Orth}) with ${\cal S} = 0$..}
\end{figure}

The fast quantum mechanical molecular dynamics scheme, Eqs.\ (\ref{EL_3ax}) and (\ref{EL_3bx}), can also be
analyzed in terms of the convergence to ``exact'' Born-Oppenheimer molecular dynamics as a function 
of the finite integration time step $\delta t$. By comparing the deviation in forces, net Mulliken charges,
and the total energy, between the fast quantum mechanical molecular dynamics scheme and an ``exact''  Born-Oppenheimer 
molecular dynamics as a function of $\delta t$ we can study the consistency between the two methods.
Figure \ref{Figure_7} shows the difference between a fully converged ``exact''  Born-Oppenheimer molecular dynamics simulation
and the fast quantum mechanical molecular dynamics scheme as measured by the root mean square deviation
over 200 fs of simulation time.  We find that the deviation of the nuclear forces, the charges $\{q_i\}$, 
as well as the total energy difference are of the order $\delta t^2$ with a small pre-factor. 
This convergence demonstrates a consistency between the fast quantum mechanical 
scheme and Born-Oppenheimer molecular dynamics using Verlet integration, 
where the optimization-free scheme behaves as a well controlled 
and tunable approximation.  As in ``exact'' Born-Oppenheimer molecular dynamics, the dominating error is governed 
by the local truncation error arising from the choice of finite
integration time step $\delta t$, which is much larger than any difference between
the fast quantum mechanical molecular dynamics and Born-Oppenheimer molecular dynamics. 

\section{Summary and Conclusions}

Based on a free energy formulation of extended Lagrangian Born-Oppenheimer molecular dynamics in 
the limit of vanishing self-consistent-field optimization, we have derived and 
demonstrated a fast quantum mechanical molecular dynamics scheme, 
Eqs.\ (\ref{EL_3ax}) and (\ref{EL_3bx}), which with a high precision can match
the accuracy and fidelity of Born-Oppenheimer molecular dynamics. In addition to the removal of
the self-consistent-field optimization we have also demonstrated compatibility with low pre-factor linear scaling solvers.
The combined scheme provides a very efficient, energy conserving, low-complexity method to perform accurate
quantum molecular dynamics simulations. Our findings show how the computational gap between
classical and quantum mechanical molecular dynamics simulations can be reduced significantly.

\section{Acknowledgment}

We acknowledge support by the United States Department of Energy Office of Basic Energy Sciences
and the LANL Laboratory Directed Research and Development Program.  
Discussions with E. Chisolm, J. Coe, T. Peery, S. Niklasson, C. Ticknor, C.J. Tymczak, and G. Zheng, 
as well as stimulating contributions at the T-Division Ten Bar Java group are gratefully acknowledged.  
LANL is operated by Los Alamos National Security, LLC, 
for the NNSA of the U.S. DOE under Contract No. DE-AC52-06NA25396.

\section{Appendix}

\subsection{Calculating the forces in Hartree-Fock theory} \label{HF_appendix}

Here we present some details of the fast quantum mechanical molecular dynamics, 
Eqs.\ (\ref{EL_3ax}) and (\ref{EL_3bx}),
using a simple but general Hartree-Fock formalism, which should be directly applicable to a broad class of hybrid 
and semi-empirical electronic structure schemes. Instead of the auxiliary density variable $n({\bf r})$ we will here use
the more general density matrix $P$. In this formalism the extended free-energy Lagrangian in Eq.\ (\ref{X}) is given by
\begin{equation}\label{X_HF} \begin{array}{l}
{\displaystyle {\cal L}({\bf R}, {\bf \dot R}; P, {\dot P})  = \frac{1}{2}\sum_k M_k{\dot R}_k^2 - {\cal U}[{\bf R};P] + T_e{\cal S}[{\bf R};P] }\\
{\displaystyle + \frac{1}{2}\mu Tr[{\dot P}^2] - \frac{1}{2} \mu \omega^2 Tr[({\mathcal D}_{\rm gs}-P)^2],}\\
\end{array}
\end{equation}
with the potential energy chosen as  
\begin{equation}\label{HF_POT}
{\cal U}[{\bf R};P] = 2Tr[hD(P)] + Tr\{D(P)G[D(P)]\},
\end{equation}
and ground state (gs) density matrix ${\mathcal D}_{\rm gs}$.
${\cal S}[{\bf R};P]$ is an unspecified electronic entropy term, which will be 
determined by the requirement to make the derived forces variationally correct, and $T_e$ is the electronic temperature.
The Fockian (or the effective single-particle Hamiltonian) is
\begin{equation}
F[P] = h + G[P],
\end{equation}
with the short-hand notation, $G[P] = 2J[P]-K[P]$, where
$J[P]$ and $K[P]$ are the conventional Coulomb and exchange matrices, and $h$
is the matrix of the one-electron part \cite{Roothaan,RMcWeeny60}.
The temperature dependent density matrix 
\begin{equation}\label{Fermi_Op}
D(P) = Z\left(e^{\beta(F^{\perp}[P] -\mu_0 I)}+1\right)^{-1}Z^T,
\end{equation}
which corresponds to $\sigma[n]$ in Eq.\ (\ref{EL_4c}), is given as a Fermi function of
the orthogonalized Fockian,
\begin{equation}
F^{\perp}[P] = Z^TF[P]Z.
\end{equation}
Here $Z$ and its transpose $Z^T$ are the inverse L\"{o}wdin or Cholesky-like factors of the overlap matrix, $S$,
determined by the relation
\begin{equation}
Z^TSZ = I.
\end{equation}
At zero electronic temperature ($T_e = 0$ K) the Fermi-operator expansion in Eq.\ (\ref{Fermi_Op}) is given by the Heaviside step function,
with the step formed at the chemical potential $\mu_0$, separating the
occupied from the unoccupied states.

The Euler-Lagrange equations of motion of ${\cal L}$ in Eq.\ (\ref{X_HF}) are given by 
\begin{equation}\begin{array}{l}
{\displaystyle M_k{\ddot R}_k = - \left.{\frac{\partial {\cal U}}{\partial R_k}}\right\rvert_{P} + T_e\left.{\frac{\partial {\cal S}}{\partial R_k}}\right\rvert_{P}}\\
~~\\
{\displaystyle -\left.{\frac{1}{2} \mu \omega^2 \frac{\partial}{\partial R_k} Tr[({\mathcal D}_{\rm gs}-P)^2]}\right \rvert_{P},}
\end{array}
\end{equation}
and
\begin{equation}
\mu {\ddot P} = \mu \omega^2 ({\mathcal D}_{\rm gs}-P) - \left.{\frac{\partial {\cal U}}{\partial P}}\right\rvert_{\bf R} + T_e\left.{\frac{\partial {\cal S}}{\partial P}}\right\rvert_{\bf R}.
\end{equation}
A cumbersome but fairly straightforward derivation (see Ref. \cite{ANiklasson08b} for a closely related example), 
using the relation and notation $Z_{R_k} = \partial Z/\partial R_k = -(1/2)S^{-1}S_{R_k}Z$,
and defining the ${\cal S}$ term such that
\begin{equation}\label{SCond_1}
 T_e\left.{\frac{\partial {\cal S}}{\partial P_{ij}}}\right\rvert_{\bf R} = 2 Tr\left[F^{\perp}[D] D^{\perp}_{P_{ij}}\right]
\end{equation}
and
\begin{equation}\label{SCond_2}
T_e\left.{\frac{\partial {\cal S}}{\partial R_k}}\right\rvert_{P} = 2 Tr\left[F^{\perp}[D] D^{\perp}_{R_k}\right],
\end{equation}
gives the equations of motion
\begin{equation}\begin{array}{l}
{\displaystyle M_k{\ddot R}_k =  -2Tr[h_{R_k}D] - Tr[DG_{R_k}(D)] }\\ 
~~\\
{\displaystyle + Tr[(DF[D]S^{-1}+ S^{-1}F[D]D)S_R]}\\
~~\\
{\displaystyle  -\left.{\frac{1}{2} \mu \omega^2 \frac{\partial}{\partial R_k} Tr[({\mathcal D}_{\rm gs}-P)^2]}\right \rvert_{P}},
\end{array}
\end{equation}
and
\begin{equation}
\mu {\ddot P} = \mu \omega^2 ({\mathcal D}_{\rm gs}-P).
\end{equation}
Notice that because of matrix symmetry $P_{ij}$ is not independent form $P_{ji}$. The partial derivatives 
of matrix elements $P_{ij}$ are therefore both over $P_{ij}$ and $P_{ji}$.
In the limit $\mu \rightarrow 0$, we get the final equations of motion for the fast quantum mechanical molecular dynamics scheme,
\begin{equation}\label{EL_1}\begin{array}{l}
{\displaystyle M_k{\ddot R}_k =  -2Tr\left[h_{R_k}D\right] - Tr\left[DG_{R_k}(D)\right]}\\
~~\\
{\displaystyle  + Tr\left[(DF[D]S^{-1} + S^{-1}F[D]D)S_{R_k}\right]},
\end{array}
\end{equation}
\begin{equation}\label{EL_2}
{\ddot P} = \omega^2 \left(D(P)-P\right),
\end{equation}
where we have included the substitution of ${\mathcal D}_{\rm gs}$ with $D(P)$ in the same way as in Eq.\ (\ref{EL_3bx}), i.e.
with $\omega^2$ rescaled by a constant $c \le 1$.
The notation for the partial derivative of the two-electron term is defined 
as $G_{R_k}(D) = \left.{\partial G(D)/\partial R_k}\right\rvert_{D}$,
i.e. under the condition of constant density matrix $D$. 

The last term in Eq.\ (\ref{EL_1}), which includes the basis-set dependence $S_{R_k}$ is the Pulay force term that
here is given in a generalized form that is valid also for non-idempotent density matrices at finite electronic 
tememperatures \cite{ANiklasson08b}.

\subsection{Approximate Entropy contribution}

The ${\cal S}[{\bf R};P]$ term is defined such that the two conditions in Eqs. (\ref{SCond_1}) and (\ref{SCond_2})
are fulfilled. At the self-consistent ground state density, i.e. when $P = D = {\mathcal D}_{\rm gs}$, 
both these conditions are automatically satisfied
by the corresponding regular ground state (gs) electronic entropy contribution to the free energy \cite{RParr89},
\begin{equation}\label{SED}\begin{array}{l}
{\displaystyle {\cal S}_{\rm gs}[{\bf R};P] = {\cal S}_{\rm gs}[{\bf R};D^\perp(P)] }\\ 
~~\\
{\displaystyle = -2 k_{\rm B} Tr[D^\perp\ln(D^\perp) + (I-D^\perp)\ln(I-D^\perp)]},
\end{array}
\end{equation}
where the relation between $D^\perp$ and $D$ is given by the congruence transformation
\begin{equation}\label{ZDZ}
D = ZD^\perp Z^T.
\end{equation}
A related derivation is given in Ref. \cite{ANiklasson08b}.
Using the approximate estimate ${\cal S}_{\rm gs}[{\bf R};P]$ in Eq.\ (\ref{SED}) when $P$ and $D$ deviate from the ground state gives,
\begin{equation}\begin{array}{l}
{\displaystyle  T_e\left.{\frac{\partial {\cal S}}{\partial P_{ij}}}\right\rvert_{\bf R} = 2 Tr\left[F^{\perp}[P] D^{\perp}_{P_{ij}}\right]},
~~\\
{\displaystyle T_e\left.{\frac{\partial {\cal S}}{\partial R_k}}\right\rvert_{P}  = 2 Tr\left[F^{\perp}[P] D^{\perp}_{R_k}\right]},\\ 
\end{array}
\end{equation}
which only approximately fulfills the conditions in Eqs. (\ref{SCond_1}) and (\ref{SCond_2}).
It is possible to show that the error is linear in $\delta P = D-P$ by a linearization of $F^{\perp}[D]$ around $P$.
Since $D(P)$ and $P$ can be assumed to be close to the ground state, $\delta P$ is small. 
From the scaling result illustrated in Fig. \ref{Figure_7} 
the error should therefore be quadratic in the integration time step, i.e.  $\sim \delta t^2$. 
We may therefore approximate the total free energy
using ${\cal S}_{\rm gs}[{\bf R};P]$, which is zero at $T_e = 0$ K. However, for the exact formulation and derivation 
of the equations of motion, Eqs.\ (\ref{EL_1}) and (\ref{EL_2}), the entropy contribution, $T_e{\cal S}[{\bf R};P]$, 
is unknown, both at finite and zero temperatures. As is seen in the equations of motion, Eqs.\ (\ref{EL_1})
and ({\ref{EL_2}), this does not affect the forces or the dynamics, only the estimate of the constant
of motion,
\begin{equation}
E_{\rm tot} = \frac{1}{2}\sum_k M_k{\dot R}_k^2 + {\cal U}[{\bf R};P] - T_e{\cal S}[{\bf R};P] ,
\end{equation}
is approximated. By comparing the approximate $E_{\rm tot}$ to optimized ``exact'' Born-Oppenheimer
molecular dynamics simulations, the accuracy of the dynamics can be estimated.

\subsection{Alternative potential energy forms}

As an alternative to the potential energy, ${\cal U}({\bf R};P)$, in Eq.\ (\ref{HF_POT}) we may chose
other functional forms that are equivalent at the ground state, i.e. when $P=D={\cal D}_{gs}$.
By using the Harris-Foulkes-like relation \cite{JHarris85,WMCFoulkes89},
\begin{equation}\label{Approx}
{\displaystyle Tr[DG(D)] \approx Tr[(2D-P)G(P)]},
\end{equation}
which has an error of second order in $\delta P = D-P$, we may, for example, choose
\begin{equation}\label{HF_POT_2}
{\cal U}[{\bf R};P] = 2Tr[hD(P)] + Tr\{[2D(P)-P]G(P)\},
\end{equation}
as our potential energy term.
In this case, the equations of motion at $T_e = 0$ corresponding to Eqs.\ (\ref{EL_1}) and (\ref{EL_2})
become
\begin{equation}\label{NucOrth}\begin{array}{l}
{\displaystyle M_k{\ddot R}_k =  -2Tr\left[h_{R_k}D\right] - Tr\{[2D-P]G_{R_k}(P)\}}\\
~~\\
{\displaystyle  + Tr\left[(DF[P]S^{-1} + S^{-1}F[P]D)S_{R_k}\right]},
\end{array}
\end{equation}
and
\begin{equation}\label{ElOrth}
{\ddot P} = \omega^2 \left(D-P\right),
\end{equation}
with the constant of motion
\begin{equation}\label{EtotOrth}\begin{array}{l}
{\displaystyle E_{\rm tot} = \frac{1}{2}\sum_k M_k{\dot R}_k^2 + 2Tr[hD]} \\
~~\\
{\displaystyle + Tr\{(2D-P)G(P)\} - T_e {\cal S}({\bf R};P)}.
\end{array}
\end{equation}
The entropy term that makes the nuclear forces variationally correct is here fulfilled by
the expression in Eq.\ (\ref{SED}). With this choice of potential our dynamics only 
requires one Fockian or effective single particle Hamitonian construction per time step.
Unfortunately, the error in the Pulay force has been found to be large compared to Eq.\ (\ref{EL_1}).
The dynamics in Eqs. (\ref{NucOrth}) and (\ref{ElOrth}) should therefore be used
only for orthogonal representations, i.e. when the overlap matrix $S = I$.

\subsection{Self-Consistent-Charge Density Functional Tight-Binding Theory} \label{sctb}

In self-consistent-charge density functional based tight-binding theory \cite{APSutton88,MFinnis98,MElstner98,MFinnis03} 
the continuous electronic density, $\sigma({\bf n})$, or the density matrix, $D(P)$, in Eq.\ (\ref{HF_POT}) is replaced by the net Mulliken charges ${\bf q}[{\bf n}] = \{q_i\}$ for each atom $i$, where ${\bf n} = \{n_i\}$ are the dynamical
variables corresponding to $P$. The potential energy functional 
${\cal U}$ in Eq.\ (\ref{HF_POT}) is then reduced to
\begin{equation}\label{DFTB_1}
{\cal U}[{\bf R};{\bf n}] = 2\sum_{i \in \rm occ} \varepsilon_i - \frac{1}{2} \sum_{i,j} q_i({\bf n})q_j({\bf n})\gamma_{ij} + E_{\rm pair}[{\bf R}].
\end{equation}
Here $\varepsilon_i$ are the (doubly) occupied eigenvalues of the charge dependent effective single-particle Hamiltonian
\begin{equation}\label{DFTB_2}
\begin{array}{l}
{\displaystyle H_{i\alpha,j\beta}[{\bf n}] = h_{i\alpha,j\beta} } \\
~~\\
{\displaystyle + (1/2)\sum_{k \beta'} \left( S_{i\alpha,k\beta'} V^{ee}_{k\beta',j\beta}
+ V^{ee}_{i\alpha,k\beta'} S_{k\beta',j\beta} \right)}
\end{array}
\end{equation}
where
\begin{equation}
 V^{ee}_{j\beta,k\beta'}  = \sum_l q_l({\bf n})\gamma_{jl}\delta_{jk}\delta_{\beta'\beta},
\end{equation}
$h_{i\alpha,j\beta}$ is a parameterized Slater-Koster tight-binding Hamiltonian,
$S_{i\alpha,j\beta}$ the overlap matrix,
$i$ and $j$ are atomic indices and $\alpha$ and $\beta$ are orbital labels \cite{ESanville10}. 
The net Mulliken charges are given by
\begin{equation}
q_i[{\bf n}]\ = 2\sum_{\alpha \in i} \left(\varrho^{\perp}_{i\alpha,i\alpha} - {\varrho^0}^{\perp}_{i\alpha,i\alpha}\right),
\end{equation}
with the density matrix
\begin{equation}\label{DM_perp}
\varrho^{\perp} = \varrho^{\perp}[{\bf n}]  = \left(e^{\beta(H^{\perp}[{\bf n}] - \mu_0 I)}+1\right)^{-1},
\end{equation}
using the orthogonalized Hamiltonian
\begin{equation}
H^{\perp}[{\bf n}]  = Z^T H[{\bf n}] Z.
\end{equation}
Here ${\varrho^0}$ is the density matrix of the corresponding separate non-interacting atoms.
The de-orthogonalized density matrix is
\begin{equation}
\varrho = \varrho[{\bf n}]  = Z\varrho^{\perp}[{\bf n}] Z^T,
\end{equation}
and as above, the congruence transformation factors are defined through
\begin{equation}\label{Congr}
Z^T S Z = I,
\end{equation}
where $S$ is the basis set overlap matrix.

The electron-electron interaction in Eq.\ (\ref{DFTB_1}) is determined by $\gamma_{ij}$, which decays like $1/R$ at large distances and equals
the Hubbard repulsion for the on-site interaction. 
$E_{\rm pair}[{\bf R}]$ is a sum of pair potentials, $\phi(R)$, that provide short-range repulsion. The radial dependence, $\zeta(R)$, 
of the Slater-Koster bond integrals, elements of the overlap matrix, and the $\phi(R)$ are all represented analytically 
in \textsc{latte} by the mathematically convenient form, 
\begin{equation}
\zeta(R) = A_0 \prod_{i=1}^4 \exp{(A_i R^i)},
\end{equation}
where $A_0$ to $A_4$ are adjustable parameters that are fitted to the results of quantum 
chemical calculations on small molecules. To ensure that the off-diagonal elements of $h$ 
and $S$ and the $\phi(R)$ in our self-consistent tight-binding implementation decay smoothly 
to zero at a specified distance, $R_\text{cut}$, we replace the $\zeta(R)$ by cut-off tails of the form, 
\begin{multline}
 t(R) = B_0 + \Delta R(B_1 + \Delta R(B_2 \\ 
+ \Delta R(B_3 + \Delta R(B_4 + \Delta R B_5 ))))
\end{multline}
at $R = R_1$, where $\Delta R = R - R_1$ and $B_0$ to $B_5$ are adjustable parameters. The adjustable 
parameters are parameterized to match the value and first and second derivatives of $t(R)$ 
and $\zeta(R)$ at $R = R_1$ and to set the value and first and second derivatives of $t(R)$ to zero at $R = R_\text{cut}$.

\subsubsection{Non-orthogonal representation at $T_e \ge 0$}

The fast quantum mechanical molecular dynamics scheme, Eqs.\ (\ref{EL_1}) and (\ref{EL_2}) or Eqs.\ (\ref{EL_3a}) and (\ref{EL_3b}), 
using self-consistent tight-binding theory in its non-orthogonal formulation is given by 
\begin{equation}\label{EL_Nonorth_a}\begin{array}{l}
{\displaystyle M_k{\ddot R}_k  = -2Tr\left[\varrho H_{R_k}\right]  }\\
~~\\
{\displaystyle + \frac{1}{2} \sum_{i,j}  q_iq_j \frac{\partial\gamma_{ij}}{\partial R_k} +   \sum_{i,j} q_i \gamma_{ij} \left.{\frac{\partial q_j}{\partial R_k}}\right\rvert_{\varrho} }\\
~~\\
{\displaystyle + Tr[(S^{-1}H[{\bf q}]\varrho + \varrho H[{\bf q}]S^{-1})S_{R_k}]
- \frac{\partial E_{\rm pair}[{\bf R}]}{\partial R_k},}\\
\end{array}
\end{equation}
and
\begin{equation}\label{EL_Nonorth_b}
{\ddot n}_i = \omega^2 \left( q_i-n_i\right),
\end{equation}
where
\begin{equation}\label{EL_Nonorth_c}
H_{R_k} = \left.{\frac{\partial H}{\partial R_k}}\right\rvert_{\varrho}
\end{equation}
and
\begin{equation}\label{EL_Nonorth_d}
S_{R_k} = \frac{\partial S}{\partial R_k}.
\end{equation}

The partial derivatives of $q_j$ and $H$ in Eqs.\ (\ref{EL_Nonorth_a}) and (\ref{EL_Nonorth_c}) are
with respect to a constant density matrix $\varrho$ in its non-orthogonal form, i.e. including an $S$ dependence of $q_j$,
\begin{equation}
\left.{\frac{\partial q_j}{\partial R_k}}\right\rvert_{\varrho} = 2 \sum_{\alpha \in j} \left( \varrho S_{R_k} \right)_{j\alpha, j\alpha}.
\end{equation}

The total energy is given by
\begin{equation}\label{Etot_NonOrth}\begin{array}{l}
{\displaystyle E_{\rm tot} = \frac{1}{2}\sum_k M_k{\dot R}_k^2 + 2\sum_{i \in \rm occ} \varepsilon_i }\\
~~\\
{\displaystyle - \frac{1}{2} \sum_{i,j} q_iq_j\gamma_{ij} + E_{\rm pair}[{\bf R}] - T_e{\cal S}[{\bf R};{\bf n}]}, 
\end{array}
\end{equation}
with the entropy contribution to the free energy approximated by
\begin{equation}\label{Ent_Approx}
{\cal S}[{\bf R};n] \approx  -2k_{\rm B} \sum_i \left\{f_i \ln(f_i) + (1-f_i)\ln(1-f_i)\right\}.
\end{equation}
Here $f_i = f_i[n]$ are the eigenstates of the Fermi operator expansion $\varrho^{\perp}[n]$ of $H^{\perp}[{\bf n}]$ in Eq.\ (\ref{DM_perp}). 

\subsubsection{Orthogonal representation at $T_e = 0$}

For orthogonal formulations, i.e. when $S=I$, and at zero electronic temperature, $T_e=0$, we will base our dynamics on the 
equations of motion in Eqs.\ (\ref{NucOrth}) and (\ref{ElOrth}).
In this case the fast quantum mechanical molecular dynamics scheme, Eqs.\ (\ref{EL_Nonorth_a})-(\ref{EL_Nonorth_b}), is given by
\begin{equation}\label{EL_Orth_a}\begin{array}{l}
{\displaystyle M_k{\ddot R}_k  = -2Tr\left[\varrho H_{R_k}\right]  + \frac{1}{2} \sum_{i,j} \left( n_in_j \frac{\partial\gamma_{ij}}{\partial R_k} \right)}\\
{\displaystyle - \frac{\partial E_{\rm pair}[{\bf R}]}{\partial R_k}},
\end{array}
\end{equation}
\begin{equation}\label{EL_Orth_b}
{\ddot n}_i = \omega^2 \left( q_i-n_i\right),
\end{equation}
where
\begin{equation}\label{EL_Orth_c}
\{H_{R_k}[{\bf n}]\}_{i\alpha,j\beta} = \frac{\partial h_{i\alpha,j\beta}}{\partial R_k} +
\sum_l n_l\frac{\partial \gamma_{il}}{\partial R_k}  \delta_{ij}\delta_{\alpha\beta}.
\end{equation}
The density matrix is given directly from the step function of the Hamiltonian, $\varrho  = \theta(\mu_0 I - H[{\bf n}])$,
without any de-orthogonalization that requires the calculation of the inverse factorization of the overlap matrix, Eq.\ (\ref{Congr}). 
The constant of motion, $E_{\rm tot}$, is approximate by
\begin{equation}\label{Etot_Orth}\begin{array}{l}
{\displaystyle E_{\rm tot} = \frac{1}{2}\sum_k M_k{\dot R}_k^2 + 2\sum_{i \in \rm occ} \varepsilon_i}\\
~~\\
{\displaystyle  - \frac{1}{2} \sum_{i,j} (2n_i-q_i)n_j\gamma_{ij} + E_{\rm pair}[{\bf R}]}.
\end{array}
\end{equation}

\subsubsection{General remarks}

Apart from the first few initial molecular dynamics time steps, where we apply a
high degree of self-consistent-field convergence and set ${\bf n = q}$, no ground state self-consistent-field optimization is required.
The density matrix, $\varrho$, and the Hamiltonian, $H$, necessary in the force calculations (and for the total energy) are calculated only 
once per time step in the orthogonal case with one additional construction of the Hamiltonian required in non-orthogonal simulations.  
The numerical integration of the equations of motion in Eq.\ (\ref{EL_3ax}) is performed 
with the velocity Verlet scheme and in Eq.\ (\ref{EL_3bx}) with the modified Verlet scheme in Eq.\ (\ref{Verl_n})
as described in Ref. \cite{ANiklasson09}. For the examples presented here we 
used the modified Verlet scheme including dissipation ($\alpha > 0$) with $K=5$ and the constant $\kappa = \delta t^2 \omega^2$ 
as given in Ref. \cite{ANiklasson09} was rescaled by a factor $1/2$ in all examples except for one of the test cases in Fig.\ \ref{halfeV}.

%~\\
%{\bf Author contributions}
%~\\
%A.M.N. proposed the fast QMMD scheme. M.C. and A.M.N. performed the implementations,
%simulations, analyzed the results and prepared the manuscript.
%
%~\\
%{\bf Additional information}
%~\\
%The authors declare no competing financial interests. 
%Correspondence should be addressed to A.M.N. (amn@lanl.gov)

\end{document}